\documentclass{elsart}

\usepackage{natbib}

 \usepackage{epsfig}

\usepackage{amssymb}

\begin{document}

\begin{frontmatter}

% Title, authors and addresses

% use the thanksref command within \title, \author or \address for footnotes;
% use the corauthref command within \author for corresponding author footnotes;
% use the ead command for the email address,
% and the form \ead[url] for the home page:
% \title{Title\thanksref{label1}}
% \thanks[label1]{}
% \author{Name\corauthref{cor1}\thanksref{label2}}
% \ead{email address}
% \ead[url]{home page}
% \thanks[label2]{}
% \corauth[cor1]{}
% \address{Address\thanksref{label3}}
% \thanks[label3]{}

\title{Large-scale solar cycle features of solar photospheric magnetic field\thanksref{footnote1}}
\thanks[footnote1]{This work is jointly supported by the National Natural Science
Foundation of China (40604019) and the 973 project under grant
2006CB806304.}

% use optional labels to link authors explicitly to addresses:
% \author[label1,label2]{}
% \address[label1]{}
% \address[label2]{}

\author{W.B. Song\corauthref{cor}}
\address{State Key
Laboratory for Space Weather, Center for Space Science and Applied
Research, Chinese Academy of Sciences, Beijing 100080, China}
\corauth[cor]{W.B. Song} \ead{wbsong@spaceweather.ac.cn}
%url can be given like this
%\ead[url]{http://authors.elsevier.com/locate/latex}

\begin{abstract}
% Text of abstract

It is well accepted that the solar cycle originates from a
magnetohydrodynamics dynamo deep inside the Sun. Many dynamo
models have long been proposed based on a lot of observational
constraints. In this paper, using 342 NSO/Kitt Peak solar synoptic
charts we study the solar cycle phases in different solar
latitudinal zones to set further constraints. Our results can be
summarized as follows. (1) The variability of solar polar regions'
area has a correlation with total unsigned magnetic flux in
advance of 5 years. (2) The high-latitude region mainly appears
unipolar in the whole solar cycle and its flux peak time lags
sunspot cycle for 3 years. (3) For the activity belt, it is not
surprised that its phase be the same as sunspot's. (4) The flux
peak time of the low-latitude region shifts forward with an
average gradient of 32.2 $day/deg$. These typical characteristics
may provide some hints for constructing an actual solar dynamo.

\end{abstract}


\end{frontmatter}

% main text
\section{Introduction}

Dynamo theory recognized to reproduce the solar cycle has
continuously been a hot topic in solar physics since fifty years
ago (e.g., Parker, 1955). For example, in the classical
$\alpha-\Omega$ turbulent dynamo, two basic processes are
involved. The first one is $\Omega$-effect which shears
pre-existing poloidal fields by differential rotation to produce a
relatively strong toroidal fields; the second one is
$\alpha$-effect which lifts and twists toroidal flux tubes to
regenerate poloidal fields (Parker, 2001). Nowadays many dynamo
models have been developed in which the major observational
constraints come from the sunspot behavior, such as butterfly
diagram, differential rotation, Hale magnetic solar cycles, and so
on. In addition, based on a deep meridional flow found in both
hemispheres, some flux-transport dynamo models are constructed
(see Dikpati, 2005). Li et al. (2005) design a flux tube dynamo
model by use of solar internal rotation on the assumption of a
downflow effect under photosphere. In this paper, using solar
magnetic synoptic charts we investigate the large-scale solar
cycle features in different latitudinal strips and expect such
information can provide helps for constructing a more reliable
dynamo.

\begin{figure}
\label{figure1}
\begin{center}
\includegraphics*[width=12cm,angle=0]{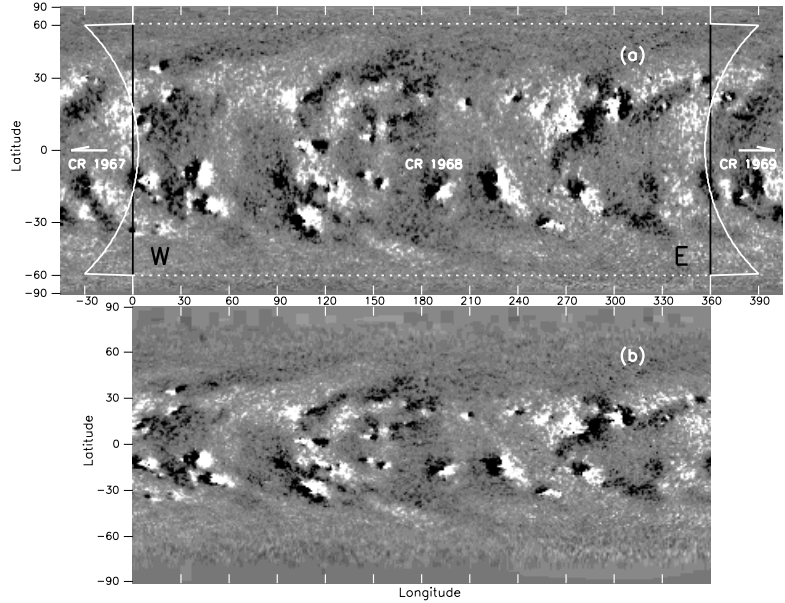}
\end{center}
\caption{(a) Three connected NSO/Kitt Peak magnetic synoptic
charts of CR1967-1969. Two solid white lines outline the real
region corresponding to full solar disk. (b) The newly constructed
synoptic chart with a grid of 360 equal steps in longitude by 180
equal steps in latitude.}
\end{figure}

\section{Database}

In this paper we use NSO/Kitt Peak magnetic synoptic charts during
Carrington Rotations (CRs) 1666-2007 as our database which show
the distribution of solar radial magnetic fields. Each magnetogram
covers one CR and has a grid of 360 equal steps in longitude by
180 equal steps in sine latitude. First, as shown in figure 1a, we
connect all charts in time order (Stenflo \& G\"{u}del 1988) and
outline the real region corresponding to full solar disk. In the
range of latitude $|\theta|\leq60^{\circ}$, two concave sidelines
are due to differential rotation (Song \& Wang 2005). Meanwhile,
for the polar regions (PRs) where $|\theta|>60^{\circ}$, we choose
the time scale to keep its original Carrington coordinate on the
consideration of most polar coronal holes having a rigid rotation.
Second, we resize each latitudinal strip to a uniform dimension of
360 elements by a linear shrinkage or interpolation method. Then
in the vertical direction we modulate the equal steps in sine
latitude to equal steps in latitude to cut down the measurement
errors near polar regions. Finally, a synoptic chart with a new
grid is constructed, see figure 1b.

\begin{figure}
\label{figure1}
\begin{center}
\includegraphics*[width=12cm,angle=0]{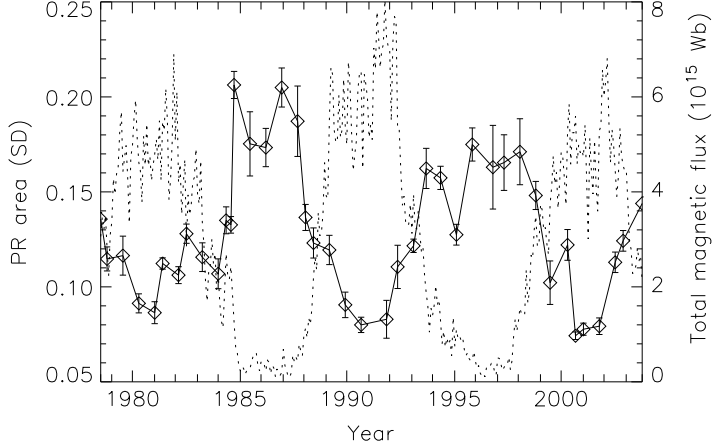}
\end{center}
\caption{PRs' area (solid) versus total unsigned magnetic flux
sequence (dashed) during solar cycles 21 to 23. 1 SD means the
area of whole solar disk.}
\end{figure}

\section{Analytical results}

In reference to the result of Song \& Wang (2006), we divide each
hemisphere into four latitudinal zones, which represent the PR,
the high-latitude region, the activity belt and the low-latitude
region.

\subsection{The polar region}

For two PRs, because of an obvious projection effect in the
measurement of magnetic field, we analyze the area instead of
total magnetic flux. In order to identify PRs' boundaries in Kitt
peak magnetograms, we set two criterions: (1) During solar
minimums, both PRs are unipolar flux regions; (2) During solar
maximums, the outer boundaries of plage regions can be regarded as
those of PRs. The measurement result is displayed in figure 2 in
which the time resolution is about 10 CRs and the average
measurement error is about 7\%. For making a good comparison we
superpose the total unsigned magnetic flux (we measure it also
using Kitt Peak charts and choose pixels with value higher than 20
Gauss in the range of $|\theta|\leq60^{\circ}$) in it. It is
obvious that PRs' area varies in completely anti-phase to the flux
sequence. From their relative peak strengths we further find a
coherence existing between the flux sequence and PRs' area with a
time lag of 5 years. Makarov et al. (2003) find a similar relation
using the duration of the polarity reversal in the octupole
component of solar magnetic field. Such a correlation in strength
may give evidence of solar cycle to be initiated from PRs just as
described by Babcock (1961).

\begin{figure}
\label{figure1}
\begin{center}
\includegraphics*[width=12cm,angle=0]{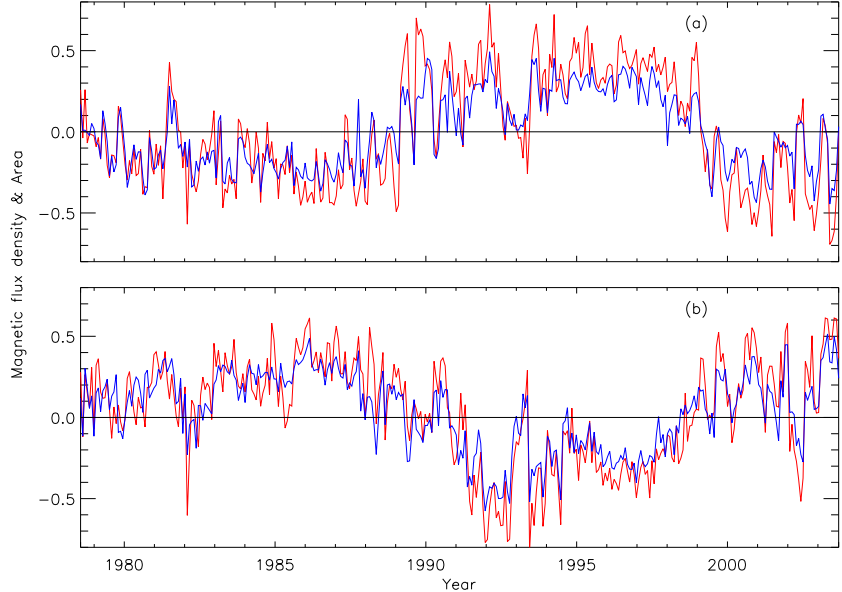}
\end{center}
\caption{The proportions of mean magnetic flux density (blue) and
area (red) between two magnetic polarities in high-latitude
regions. Panel $a$ is for the northern hemisphere, panel $b$ is
for the southern hemisphere.}
\end{figure}

\subsection{The high-latitude region}

The borderline between the high-latitude region and the activity
belt is chosen the position where $\theta=\pm40^{\circ}$. However
sometimes we can meet several active regions located higher than
this latitude, then the borderline will be adjusted to be a bit
higher than these active regions. We mainly investigate two
parameters in this region. They are $m=\frac{m_p-m_n}{m_p+m_n}$
and $a=\frac{a_p-a_n}{a_p+a_n}$, where $m$ indicates the mean
magnetic flux density, $a$ indicates the area, $p$ is for the
positive magnetic flux and $n$ is for the negative magnetic flux.
The result is shown in figure 3. From it we can find such two
parameters vary simultaneously and the high-latitude region always
appear unipolar (the dominant polarity occupies more than 70\%)
except during a short time interval for the polarity reversal
every other solar cycle. It is notable that the epochs when the
polarity signs change is near the years of 1979, 1989 and 1999
which lag solar minimums for about 3 years. The main magnetic
feature in high-latitude regions is plage which is usually and
approximately coincide with faculae in the underlying photosphere.
Sheeley (1991) finds a $90^{\circ}$ phase shift between the number
of polar faculae (it would be in a more extended polar region than
ours defined in section 3.1) and the sunspot number, also with the
sunspot number occurring earlier.

\subsection{the activity belt and the low-latitude region}

Figure 4 depicts the contours of average positive and negative
magnetic flux density of all latitudinal strips (the strip width
is $1^\circ$) refer to the range of $|\theta|\leq40^\circ$. A
similar chart without the consideration of differential rotation
is drawn by Dikpati et al. (2004). We find it is very like the
butterfly diagram of sunspots. In order to analyze the flux peak
time (or the main phase of solar cycle) in different latitudinal
strips, we compute their Gaussian fits for every solar cycles. The
Gaussian centers are described by six solid curves in figure 4
from which we can find that the variabilities of flux peak time
exhibit two obvious inflexions near $\theta=\pm20^{\circ}$. In the
higher latitude regions (or two activity belts) the flux peak time
appears a bit steady which shows a good coherence with solar
maximums. However, in the low-latitude regions the flux peak time
starts to shift forward with an average gradient of $32.2\pm12.4$
$day/deg$.

\begin{figure}
\label{figure1}
\begin{center}
\includegraphics*[width=12cm,angle=0]{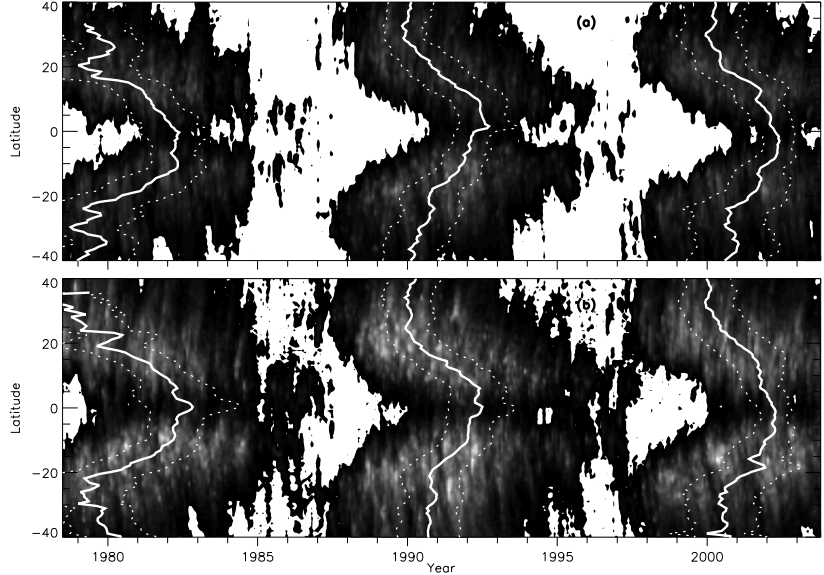}
\end{center}
\caption{Contours of average positive (a) and negative (b)
magnetic flux density refer to the range of
$|\theta|\leq40^\circ$. Six solid curves indicate the center
positions of Gaussian fit and their side dashed lines indicate
$\pm1\sigma$ uncertainty estimates.}
\end{figure}

\section{Conclusion}

According to the behaviors shown above, we find four typical solar
latitudinal strips have different solar cycle phases. (1) The
variability of PRs' area has a correlation with total unsigned
magnetic flux in advance of 5 years. (2) The high-latitude region
mainly appears unipolar in the whole solar cycle and its flux peak
time lags sunspot cycle for 3 years. (3) For the activity belt, it
is not surprised that its phase is the same as sunspot's. (4) The
flux peak time of the low-latitude region shifts forward with an
average gradient of 32.2 $day/deg$. Generally speaking, dynamo
theory is developed for reproducing the well-regulated solar
cycle. Here we propose the typical large-scale solar cycle
features of solar magnetic field and they appear to be able to
give some helps in constructing a more reliable solar dynamo.

% The Appendices part is started with the command \appendix;
% appendix sections are then done as normal sections
% \appendix

\end{document}